# Whole-Volume Clustering of Time Series Data from Zebrafish Brain Calcium Images via Mixture Modeling

Hien D. Nguyen*, Jeremy F. P. Ullmann, Geoffrey J. McLachlan, Venkatakaushik Voleti, Wenze Li, Elizabeth M. C. Hillman, David C. Reutens, and Andrew L. Janke

February 27, 2017

**Abstract**

Calcium is a ubiquitous messenger in neural signaling events. An increasing number of techniques are enabling visualization of neurological activity in animal models via luminescent proteins that bind to calcium ions. These techniques generate large volumes of spatially correlated time series. A model-based functional data analysis methodology via Gaussian mixtures is suggested for the clustering of data from such visualizations is proposed. The methodology is theoretically justified and a computationally efficient approach to estimation is suggested. An example analysis of a zebrafish imaging experiment is presented.

*Hien Nguyen and Geoffrey McLachlan are with the School of Mathematics and Physics, The University of Queensland, St. Lucia, Brisbane, Australia 4075. Hien Nguyen, David Reutens, and Andrew Janke are with the Centre for Advanced Imaging, The University of Queensland, St. Lucia, Brisbane, Australia 4075. Jeremy Ullmann is with the Department of Neurology, Boston Children's Hospital and Harvard Medical School, Boston, USA 02115. Venkatakaushik Voleti, Wenze Li, and Elizabeth Hillman are with the Laboratory for Functional Optical Imaging, Departments of Biomedical Engineering and Radiology, Columbia University, New York, New York 10027, USA. Corresponding Author: Hien Nguyen (Email: h.nguyen7@uq.edu.au).

# 1 Introduction

Calcium is a ubiquitous secondary messenger that regulates vital signaling events in neurons and other excitable cells [1]. In neurons, synaptic input and action potential firing trigger the rapid input of large quantities of calcium ions [2]. By fusing a calcium-binding domain to a fluorescent protein it is possible to monitor calcium ions and therefore various spike features [3]. Genetically encoded calcium indicators (GECIs) can target specific cell types or subcellular compartments and have facilitated long-term optical recording of genetically targeted neurons in vivo. GECIs like GCaMP are the most widely used calcium indicators [4] and a range of mouse, drosophila, and zebrafish lines have been generated that express GCaMP pan-neuronally.

In recent years, the zebrafish has become a popular model in neurological research (e.g. [5] and [6]) and the number of studies imaging neuronal calcium dynamics in larval zebrafish has increased dramatically (e.g. [7] and [8]). Larval zebrafish are particularly amenable to calcium imaging as they are translucent making it easy to image large populations of neurons without surgery, possess a relatively small brain ($\approx 800 \times 600 \times 200$ $\mu$m$^3$) thereby facilitating whole-brain imaging [9], and the generation of transgenic lines is relatively straightforward promoting the creation of a range of transgenic lines [10]. Light-sheet based microscopy methods can enable imaging in larval zebrafish with sufficient temporal resolution to record neural dynamics throughout the brain [5].

In this article, we devise a method for clustering time series that are obtained from whole-volume calcium imaging experiments. Large time series data sets, consisting of thousands of time points, were acquired using using swept confocally-aligned planar excitation microscopy (SCAPE; [11]). SCAPE is a novel method for high-speed light sheet imaging, which enables imaging of the entire zebrafish brain using a single objective lens in a flexible geometry and at very high spatiotemporal resolution. The data set utilized here was a whole-volume calcium image of a larval zebrafish brain at a spatial resolution of $640 \times 130 \times 300$ ($x \times y \times z$) voxels and a temporal resolution of 4 volumes per second.

Two primary factors make the clustering of whole-volume calcium imaging time series difficult. Firstly, the time series can often be obtained at sparse or irregular intervals. Secondly, the size of the data firmly places the problem within the realms of Big Data, where conventional methods are infeasible to implement (cf. [12, Sec. 1.2]). The method that we develop in this article is designed to address these primary concerns.

Time series clustering is a well-studied area in data analysis and signals processing; see [13] and [14] for recent literature reviews. From [14], it is clear that there are numerous approaches to the problem of time series clustering. Due to the high dimensionality of the raw time series that arise from calcium imaging experiments, an abstraction is necessary to limit computational burden. The general framework of functional data analysis (FDA) is ammenable to our aims; see [15] for an introduction to FDA. In this article, we present a mixture model-based functional data approach, thus only a brief literature review of such methods will be provided in the sequel.

In [16], a mixture of polynomial basis-filtered models was suggested for the clustering of time-dependent trajectories. In [17], a regularized mixture of mixed-effects (MEM) basis-filtered models using B-splines was considered for the clustering of sparsely-sampled univariate functional data. Both [18] and [19] suggested MEM basis-filtered models using Fourier bases for clustering of time-course gene expression data; [20] extended the method of [19] by allowing for autoregressive (AR) errors. An extension of the MEM model of [19] that allowed for concomitant covariates was proposed in [21]. An adaptation of the MEM model from [17] was suggested for the clustering of sparsely-sampled bivariate functional data in [22]. Lastly, a set of methods that are related to the mixture model-based FDA approach are the mixtures of AR models that utilize a parametric stochastic process abstraction of the time series instead; see [23], [24], and [25].

In this article, we consider a two-stage approach to clustering the calcium image obtained time series that is adapted from that of [26]. In the first stage, each time series is filtered via a common set of B-spline bases using ordinary least squares (OLS); this produces a set of basis coefficients for each time series. In the second stage, a GMM-like (Gaussian mixture model; for example, see [27, Ch. 3]) clustering of the basis coefficients for each time series is conducted to partition the time series into $k$ classes, where $k$ is determined by a BIC-like (Bayesian information criterion [28]) approach via the slope heuristic technique of [29]. In this article, we use the trimmed $k$-means algorithm [30] for clustering due to its speed in the large data context. The two-stage nature of our method shares similarities with the methods of [31] and [32].

To justify our two-stage approach, we demonstrate that it permits the same statistical interpretation as the MEM model of [17]. Although the method of [17] permits sparseness and irregularity in the time series, the iterative and slow-converging EM (expectation–maximization [33]) algorithm that is required for its implementation is inappropriate for data sizes in calcium imaging time series clustering; the same reasoning applies to the infeasibility of the MEM model-based methods; see [26, Sec. 4] for details. The methods that are based on mixtures of AR models are also inappropriate in this context due to their inability to handle sparsity in the sampling of the time series.

Along with the statistical model for calcium imaging, time series clustering, and two-stage method for performing the clustering, we also discuss the theoretical conditions under which the maximum likelihood (ML) estimator of a GMM model is consistent, under dependence between spatially correlated voxels. We then extend this discussion towards the suitability of the trimmed $k$-means algorithm. A numerical study is provided to justify our approach. To demonstrate an application of our methodology, we conduct a clustering of time series data arising from a whole-volume calcium imaging experiment of a larval zebrafish brain at rest, acquired using SCAPE microscopy.

The rest of the article proceeds as follows. A statistical model for calcium imaging time series clustering is presented in Section 2. A two-stage algorithm is suggested for performing clustering in Section 3. An application to a zebrafish brain calcium imaging data is presented in Section 4. Conclusions are drawn in Section 5. A description of the trimmed $k$-means method is also provided in the Appendix.

## 2 Statistical Model

Suppose that we wish to cluster a sample of time series $\mathcal{Y}_n = \{Y_1(t), ..., Y_n(t)\}$ that are treated as functions in $t$ on some interval $\mathbb{T}$. Let each time series $Y_i(t)$ $(i = 1, ..., n)$ be observed at $m_i$ time points $t_1, ..., t_{m_i} \in \mathbb{T}$ (thay may be sparsely and irregularly sampled) and suppose that we observe $Y_i$ at $t_j$ $(j = 1, ..., m_i)$ indirectly via the noise-corrupted variable

$$Z_i(t_j) = Y_i(t_j) + E_i(t_j), \tag{1}$$

where $E_i(t_j)$ is a random error from a univariate Gaussian density distribution with mean 0 and variance $\sigma^2$.

In order to cluster the data, we require an implicit model for the subpopulation heterogeneity of the elements in $\mathcal{Y}_n$. We propose a parametric model for the purpose. Let $Y_i(t) = \boldsymbol{B}_i^T \boldsymbol{x}(t)$, where $\boldsymbol{x}(t)$ is a $d$-dimensional vector of evaluates at point $t$ of a B-spline system with $(d-2)$-breakpoints over $\mathbb{T}$ (see [34, Ch. 9]), and let $\boldsymbol{B}_i \in \mathbb{R}^d$ be a random variable that captures the heterogeneity of $\mathcal{Y}_n$. Here, the superscript $T$ indicates matrix transposition. Cubic B-splines are used in all of our applications. We suppose that the B-spline representation $\boldsymbol{B}_i$ arises from a $k$-component GMM with density function

$$f(\boldsymbol{b};\boldsymbol{\theta}) = \sum_{c=1}^{k} \pi_c \phi_d(\boldsymbol{b};\boldsymbol{\mu}_c, \mathbf{V}_c), \tag{2}$$

where $\pi_c > 0$, $\sum_{c=1}^{k} \pi_c = 1$, $\boldsymbol{\mu}_c \in \mathbb{R}^d$, $\mathbf{V}_c \in \mathbb{R}^{d\times d}$ is positive definite ($c = 1, ..., k$), and

$$\phi_d(\boldsymbol{b};\boldsymbol{\mu}, \mathbf{V}) = |2\pi\mathbf{V}|^{-1/2} \exp\left[-\frac{1}{2}(\boldsymbol{b}-\boldsymbol{\mu})^T \mathbf{V}^{-1}(\boldsymbol{b}-\boldsymbol{\mu})\right]$$

is the multivariate Gaussian density function with mean $\boldsymbol{\mu}$ and covariance matrix $\mathbf{V}$. We shall refer to $\pi_c$ as the prior probability of observing $Y_i$ is in cluster $c$, and $\phi_d(\boldsymbol{b};\boldsymbol{\mu}, \mathbf{V})$ as the component density function of cluster $c$. Here, we put the parameter components $\pi_c$, $\boldsymbol{\mu}_c$, and $\mathbf{V}_c$ into the vector $\boldsymbol{\theta}$.

*Remark* 1. Any linear basis system can be used in place of the B-splines in this article. Examples of other basis systems are Fourier or polynomial bases; see [15, Sec. 3.3].

The model described so far is close to that of [17]. However, in order to proceed [17] estimation of each $\boldsymbol{B}_i$ is required using the data $Z_i(t_j)$, for all $i$ and $j$, via an EM algorithm. This is not feasible in the setting that we wish to conduct clustering. We instead suggest that we estimate each $\boldsymbol{B}_i$ only via the data $Z_i(t_j)$ for the same $i$. We do this via the OLS estimator

$$\tilde{\boldsymbol{B}}_i = \left(\mathbf{X}_i^T\mathbf{X}_i\right)^{-1}\mathbf{X}_i^T\boldsymbol{Z}_i, \tag{3}$$

where we let $\mathbf{X}_i^T = \left[\begin{array}{ccc} \boldsymbol{x}(t_1) & \cdots & \boldsymbol{x}(t_{m_i}) \end{array}\right]$ and $\boldsymbol{Z}_i^T = (Z_i(t_1), ..., Z_i(t_{m_i}))$. The following result is proved in [26].

**Proposition 1.** *Under characterizations (1) and (2), $\tilde{\boldsymbol{B}}_i$ (as characterized by (3)) has density function*

$$f\left(\tilde{\boldsymbol{b}}_i;\boldsymbol{\vartheta}\right) = \sum_{c=1}^{k}\pi_c\phi_d\left(\tilde{\boldsymbol{b}}_i;\boldsymbol{\mu}_c,\mathbf{V}_c+\sigma^2\left[\mathbf{X}_i^T\mathbf{X}_i\right]^{-1}\right), \tag{4}$$

*where we put the parameter components $\pi_c$, $\boldsymbol{\mu}_c$, $\mathbf{V}_c$, and $\sigma^2$ into $\boldsymbol{\vartheta}$.*

*Remark* 2. Density (4) is dependent on $i$ only via the term $\sigma^2\left[\mathbf{X}_i^T\mathbf{X}_i\right]^{-1}$. If we let $\mathbf{X}_i = \mathbf{X}$ for all $i$, then we can write

$$f\left(\tilde{\boldsymbol{b}};\boldsymbol{\psi}\right) = \sum_{c=1}^{k}\pi_c\phi_d\left(\tilde{\boldsymbol{b}};\boldsymbol{\mu}_c,\tilde{\mathbf{V}}_c\right), \tag{5}$$

where $\tilde{\mathbf{V}}_c = \mathbf{V}_c + \sigma^2\left[\mathbf{X}^T\mathbf{X}\right]^{-1}$ and $\boldsymbol{\psi}$ contains the parameter components of the model.

*Remark* 3. Alternatively to Remark 2, if we assume some structure in the sampling of the time points $t_1, ..., t_{m_i}$ for each $i$, then we can reasonably assume that $\mathbf{X}_i^T\mathbf{X}_i$ approaches some positive and invertible matrix $\boldsymbol{\Delta}$ in probability, for each $i$, as $n$ approaches infinity. In such a case, an appopriate Slutsky-type theorem would imply that $\tilde{\boldsymbol{B}}_i$ approachs a random variable with a mixture distribution of form (4), where $\mathbf{X}_i^T\mathbf{X}_i$ is replaced by $\boldsymbol{\Delta}$.

*Remark* 4. It is possible that for some choice of $m_i$ and $d$, the matrix product

$\mathbf{X}_i^T\mathbf{X}_i$ may be singular. In such situations, we may replace $\left(\mathbf{X}_i^T\mathbf{X}_i\right)^{-1}$ is (3)–(5) by the Moore-Penrose inverse $\left(\mathbf{X}_i^T\mathbf{X}_i\right)^{+}$, instead. Since the Moore-Penrose inverse is unique and always exists, its use in our application causes no unintended effects (cf. [35, Sec. 7.4]).

## 2.1 Parameter Estimation

Suppose that we observe data $\boldsymbol{z}_i$ that are realizations of $\boldsymbol{Z}_i$ from the process (1). Also suppose that $\{t_1, ..., t_{m_i}\} = \{t_1, ..., t_m\}$ for all $i$ (i.e. $\mathbf{X}_i = \mathbf{X}$ for all $i$, from Remark 2). A natural parameter estimator for $\boldsymbol{\psi}$ in (5) is the MLE (maximum likelihood estimator). Define $\nabla$ as the gradient operator, and let the MLE be $\hat{\boldsymbol{\psi}}_n$, which is a suitable root of $\ell_n(\boldsymbol{\psi}) = \sum_{i=1}^n \log f\left(\tilde{\boldsymbol{b}}_i; \boldsymbol{\psi}\right)$, where $\tilde{\boldsymbol{b}}_i = \left(\mathbf{X}^T\mathbf{X}\right)^{-1}\mathbf{X}^T\boldsymbol{z}_i$. The following useful result regarding the consistency of MLE under dependence (e.g. spatial dependence between voxels in imaging data) is obtained in [26].

**Proposition 2.** *Let $\boldsymbol{Z}_1, ..., \boldsymbol{Z}_n$ be an identically distributed (ID) and stationary ergodic (or strong-mixing) random sample (as characterized by (1)), which generates a set of OLS estimators $\tilde{\boldsymbol{B}}_1, ..., \tilde{\boldsymbol{B}}_n$. Suppose that each $\tilde{\boldsymbol{B}}_i$ ($i = 1, ..., n$) arises from a distribution with density $f\left(\tilde{\boldsymbol{b}}; \boldsymbol{\psi}_0\right)$, where $\boldsymbol{\psi}_0$ (containing $\pi_{0c}$, $\boldsymbol{\mu}_{0c}$, and $\mathbf{V}_{0c}$ for all $c$) is a strict-local maximizer of $\mathbb{E}\log f\left(\tilde{\boldsymbol{b}}; \boldsymbol{\psi}_0\right)$. If $\Psi_n = \{\boldsymbol{\psi} : \nabla\ell_n = \mathbf{0}\}$ (where we take $\Psi_n = \{\bar{\boldsymbol{\psi}}\}$, for some $\bar{\boldsymbol{\psi}}$ in the domain of $f\left(\tilde{\boldsymbol{b}}; \boldsymbol{\psi}\right)$, if $\nabla\ell_n = \mathbf{0}$ has no solution), then for any $\epsilon > 0$,*

$$\lim_{n\to\infty} \mathbb{P}\left[\inf_{\boldsymbol{\psi}\in\Psi_n} \left(\boldsymbol{\psi} - \boldsymbol{\psi}_0\right)^T \left(\boldsymbol{\psi} - \boldsymbol{\psi}_0\right) > \epsilon\right] = 0.$$

The result establishes the fact that there exists a consistent root to the likelihood score equation $\nabla\ell_n = \mathbf{0}$. This is a useful result since the likelihood for a GMM cannot possess a unique global maximum as it is unbounded and lacks identifiability. The proof of Proposition 2 invokes an extremum estimator theorem [36, Thm. 4.1.2] in conjunction with either an ergodic continuous mapping theorem or a strong-mixing continuous mapping theorem (see [37, Thms. 3.35 and 3.49]). A generic uniform law of large numbers (ULLN) such as [38, Thm. 5] can then be used to obtain the desired result; see [24, Appendix I] for the proof of a similar result.

*Remark* 5. Strong mixing can be guaranteed by assuming that the sequence $\boldsymbol{Z}_i$ is $M$-dependent (cf. [39, Sec. 2.1]). That is, there exists some $M < \infty$ such that if $|i - i'| > M$ then $\boldsymbol{Z}_i$ and $\boldsymbol{Z}_{i'}$ are independent. This is a reasonable assumption in imaging applications where features behave in locally coherent groups.

*Remark* 6. The intended use of this article is for clustering time series at voxels on a three-dimensional array. The asymptotics of Proposition 2 in conjunction with Remark 5 applies when one of the three dimensions are extended infinitely. In practice, this is unimportant, but a theoretical labelling of the voxels that would allow the $M$-dependent assumption to hold when all three dimensions are extended would be difficult. One could instead treat the problem as an estimation problem over data arising from a stationary three-dimensional mixing random field. In such a case, the $M$-dependent assumption of Remark 5 can still be made (cf. 40). The generic ULLN of [41] can be used in place of [38, Thm. 4] to establish an equivalent result.

## 2.2 Clustering via the GMM

Under (5) and following the approach from [27, Secs. 1.15.1 and 1.15.2], we have that

$$a_i = \arg \max_{c=1,\dots,k} \pi_{0c} \phi_d \left( \tilde{\boldsymbol{b}}_i ; \boldsymbol{\mu}_{0c}, \tilde{\mathbf{V}}_{0c} \right) / f \left( \tilde{\boldsymbol{b}}_i ; \boldsymbol{\psi}_0 \right) \tag{6}$$

is an outright assignment of the $i$th observation according to the optimal (Bayes) rule of allocation. Since $\boldsymbol{\psi}_0$ is unknown, we can substitute $\hat{\boldsymbol{\psi}}_n$ in (6) to obtain a plug-in allocation rule.

*Remark* 7. Extending upon Remark 2, if we also make the assumption $\pi_c = 1/k$ and $\tilde{\mathbf{V}}_c = \lambda \mathbf{I}_d$ for all $c$ and some $\lambda > 0$, then the clustering obtained from a $k$-means algorithm (e.g. [42]) approximates a clustering obtained via a GMM. The invariance of the clustering rule to the scaling parameter $\lambda$ can be seen by inspecting the discriminant function between any two classes [43, Eqn. 3.3.7]. The similarities between the algorithms for conducting GMM and $k$-means clustering are highlighted in [44, Secs. 16.1.1 and 16.1.2].

## 2.3 Model Selection

Thus far, we have not commented on the selection of the number of components $k$ as it cannot be conducted within the likelihood approach of Section 2.1. An external wrapper-type method for selecting $k$ is the penalized-contrast approach of [45]. In the context of this article, the method can be presented as follows.

Let $\mathbb{K} = \{k_1, ..., k_K\}$ be a set of $K$ possible values for the number of components and let $\hat{\boldsymbol{\psi}}_n^{[k]}$ be the MLE of (5) with $k \in \mathbb{K}$ components. Let $k_0 \in \mathbb{K}$ be the optimal number of components in the sense that it minimizes the expectation of the loss $n^{-1}\ell_n\left(\boldsymbol{\psi}_0^{[k_0]}\right) - n^{-1}\ell_n\left(\hat{\boldsymbol{\psi}}_n^{[k]}\right)$ for $k \in \mathbb{K}$. We can estimate $k_0$ by

$$\hat{k} = \arg\min_{k\in\mathbb{K}} \ -\frac{1}{n}\ell_n\left(\hat{\boldsymbol{\psi}}_n^{[k]}\right) + 2\tilde{\kappa}\ \text{pen}\left(k\right), \tag{7}$$

where $\text{pen}\left(k\right)$ is a problem-specific penalization function for a model with $k$ components and $\tilde{\kappa}$ is the so-called slope heuristic that is estimated from the data (cf. [46]) . According to [47, Tab. 1], an appropriate penalty function for the GMM is to use the number of parameters. That is, under the restricts of Remark 2, $\text{pen}\left(k\right) = \left(d^2/2 + 3d/2 + 1\right)k - 1$. Furthermore, under the additional restrictions of Remark 7, we get $\text{pen}(k) = dk$. The slope heuristic $\tilde{\kappa}$ can be estimated via the DDSE (data-driven slope estimation) method of [47]; see also [29].

# 3 Two-stage Algorithm

Make the assumptions from Remark 2. Let $\boldsymbol{z}_1, ..., \boldsymbol{z}_n$ be a realization of the random sample of time series $\boldsymbol{Z}_1, ..., \boldsymbol{Z}_n$ that are observed at the $n$ voxels, where $\boldsymbol{Z}_i^T = (Z_i(t_1), ..., Z_i(t_m))$ and $t_j \in \mathbb{T}$ ($i = 1, ..., n$ and $j = 1, ..., m$). The first stage of the algorithm is to filter each $\boldsymbol{Z}_i$ via a B-spline system. The second stage of the algorithm is to cluster the OLS estimators that are obtained upon B-spline filtering.

## 3.1 Stage 1: B-spline Filter

As in Section 2, let $\boldsymbol{x}(t)$ be a $d$-dimensional vector of evaluates at point $t$ of a B-spline system with $(d-2)$-breakpoints over $\mathbb{T}$. In this article, we find that $d = 200$ is sufficient for our application. Further, let $\mathbf{X}^T = \begin{bmatrix} \boldsymbol{x}(t_1) & \cdots & \boldsymbol{x}(t_m) \end{bmatrix}$ be the matrix of B-spline system evaluates at the $m$ sampled time points. For each $i$, we filter the time series $\boldsymbol{z}_i$ via the OLS estimator $\tilde{\boldsymbol{b}}_i = \left(\mathbf{X}^T\mathbf{X}\right)^{-1}\mathbf{X}^T\boldsymbol{z}_i$.

## 3.2 Stage 2: Clustering

Upon obtaining the OLS estimates, we now proceed to cluster the sample $\tilde{\boldsymbol{b}}_1, ..., \tilde{\boldsymbol{b}}_n$. Unfortunately, even with a modest size whole-volume brain calcium imaging dataset (e.g. $n \approx 10^6$ and $m \approx 2000$), estimating the MLE of (5) for use with Rule (6) is prohibitively computationally intensive for sufficiently large $k$ (e.g. $k \geq 10$). Even the additional restrictions imposed in Remark 7 and the use of efficient $k$-means algorithm implementations do not improve the computational speed to an acceptable level.

Making the assumptions from Remark 7, we can sufficiently-quickly estimate the component means $\boldsymbol{\mu}_c$ for $c = 1, ..., k$ (we will call the estimates $\tilde{\boldsymbol{\mu}}_{nc}$), for relatively large $k$ (e.g. $k \geq 20$) via the trimmed $k$-means method of [30] as implemented via the TCLUST algorithm of [48]. A description of the trimmed $k$-means approach is provided in the Appendix.

Upon obtaining the estimates of all the component means $\tilde{\boldsymbol{\gamma}}_n^T = (\tilde{\boldsymbol{\mu}}_{n1}, ..., \tilde{\boldsymbol{\mu}}_{nk})$, we can use Rule (6) to allocate each OLS estimate $\tilde{\boldsymbol{b}}_i$. Under the assumptions from Remark 7 and upon substitution of the component means, Rule (6) simplifies to the usual $k$-means rule

$$a_i = \arg \min_{c=1,...,k} \left(\tilde{\boldsymbol{b}}_i - \tilde{\boldsymbol{\mu}}_{nc}\right)^T \left(\tilde{\boldsymbol{b}}_i - \tilde{\boldsymbol{\mu}}_{nc}\right). \tag{8}$$

The value of $k$ is determined by Rule (7) upon estimation of $\tilde{\kappa}$ via a sufficiently large sample from $\mathbb{K} = \{2, 3, ...\}$. Here, the log-likelihood $\ell_n(\boldsymbol{\psi})$ (from Section 2.1) simplifies to

$$\ell_n(\boldsymbol{\gamma}) = \sum_{i=1}^{n} \log \sum_{c=1}^{k} k^{-1} \phi_d\left(\tilde{\boldsymbol{b}}_i; \boldsymbol{\mu}_c, \lambda \mathbf{I}_d\right) \tag{9}$$

for some fixed $\lambda > 0$. In this article, we set $\lambda = 1$.

### 3.3 Numerical Study

We perform a pair of simulation studies, S1 and S2, in order to justify the application of the assumptions from Remark 7, as well as the application of the trimmed $k$-means method for accelerated estimation. In S1, we simulate $n \in \{500, 1000, 2500, 5000\}$ curves at $m \in \{100, 200, 500, 1000\}$ uniformly-spaced points on the unit interval $\mathbb{T} = [0, 1]$. Each of the curves are sampled with equal probability from $k = 5$ classes of $(10 - 2)$-node B-spline systems that are defined by (1) and (2) and the parameter components $\boldsymbol{\mu}_1^T = (0, ..., 0)$, $\boldsymbol{\mu}_2^T = (1, 1, ..., 0)$, $\boldsymbol{\mu}_3^T = (-1, -1, ..., 0)$, $\boldsymbol{\mu}_4^T = (0, ..., 1, 1)$, $\boldsymbol{\mu}_5^T = (0, ..., -1, -1)$, $\sigma^2 = 0.25^2$, and $\mathbf{V}_c = \text{diag}\left(0.25^2, ..., 0.25^2\right)$, for each $c = 1, ..., k$. In S2, we perform the same simulation except we set

$$\mathbf{V}_c = \begin{bmatrix} 0.25^2 & 0.15^2 & \cdots & 0.15^2 \\ 0.15^2 & 0.25^2 & \cdots & 0.15^2 \\ \vdots & \vdots & \ddots & \vdots \\ 0.15^2 & 0.15^2 & \cdots & 0.25^2 \end{bmatrix}$$

for each $c = 1, ..., k$, instead.

For each combination of $m$ and $n$ in each of S1 and S2, we perform clustering using Rule (6) (i.e. GMM-based clustering without application of the assumptions of Remark 7), Rule (8) using the entire data set for the estimation of the component means, as well as Rule (8) using the trimmed $k$-means algorithm for mean estimation, with $\alpha$ set to 0.25 and 0.5. Each combination and scenario is simulated 50 times and the performance of each algorithm is measured using the adjusted-Rand index (ARI) of [49]. An ARI measure of 1 indicates a perfect match between the clustering and the generative labels, and a value of 0 indicates no association between the clustering and the generative labeling. The simulation results for S1 and S2 are report in Tables 1 and 2 as means and standard errors (SEs) of the ARI results over the 50 repetitions.

From Tables 1 and 2, we can draw the conclusion that all clustering methods appear to improve in performance with increases in $m$ and $n$. This is natural as more data allows for better estimation of the mean functions of the generative models and thus better clustering around those mean functions. In S1, we surprisingly observe that there is no tradeoff in performance between the $k$-means and GMM clustering results. Furthermore, using smaller parts of the data set (i.e. $\alpha$ is increased) appears to simply increase the variance of the ARI rather than decrease the average performance by any significant amount. We can thus conclude that when the B-spline bases are not correlated, there appears to be little loss in performance from using a trimmed $k$-means approach when compared to a full data $k$-means or a GMM clustering.

Table 1: Average time and ARI results for clustering rules applied to S1. The columns denoted $\alpha = 0.25$ and $\alpha = 0.5$ contain the results for the trimmed $k$-means estimated clusterings. ARI results are reported as means over 50 repetitions.

| | | GMM | | $k$-means | | $\alpha = 0.25$ | | $\alpha = 0.5$ | |
|---|---|---|---|---|---|---|---|---|---|
| $m$ | $n$ | ARI | SE | ARI | SE | ARI | SE | ARI | SE |
| 100 | 500 | 0.955 | 0.0024 | 0.972 | 0.0017 | 0.954 | 0.0082 | 0.965 | 0.0059 |
| 100 | 1000 | 0.969 | 0.0011 | 0.972 | 0.0010 | 0.960 | 0.0069 | 0.970 | 0.0023 |
| 100 | 2500 | 0.972 | 0.0005 | 0.971 | 0.0007 | 0.963 | 0.0063 | 0.971 | 0.0007 |
| 100 | 5000 | 0.974 | 0.0006 | 0.972 | 0.0006 | 0.949 | 0.0096 | 0.972 | 0.0006 |
| 200 | 500 | 0.970 | 0.0017 | 0.981 | 0.0012 | 0.964 | 0.0080 | 0.975 | 0.0065 |
| 200 | 1000 | 0.978 | 0.0011 | 0.982 | 0.0008 | 0.965 | 0.0072 | 0.980 | 0.0013 |
| 200 | 2500 | 0.981 | 0.0007 | 0.982 | 0.0006 | 0.971 | 0.0052 | 0.982 | 0.0005 |
| 200 | 5000 | 0.982 | 0.0004 | 0.982 | 0.0004 | 0.951 | 0.0108 | 0.982 | 0.0004 |
| 500 | 500 | 0.979 | 0.0015 | 0.986 | 0.0011 | 0.984 | 0.0016 | 0.963 | 0.0104 |
| 500 | 1000 | 0.983 | 0.0009 | 0.987 | 0.0009 | 0.970 | 0.0083 | 0.983 | 0.0041 |
| 500 | 2500 | 0.986 | 0.0006 | 0.987 | 0.0006 | 0.975 | 0.0070 | 0.987 | 0.0006 |
| 500 | 5000 | 0.987 | 0.0003 | 0.988 | 0.0003 | 0.980 | 0.0055 | 0.988 | 0.0003 |
| 1000 | 500 | 0.981 | 0.0013 | 0.986 | 0.0009 | 0.979 | 0.0040 | 0.986 | 0.0012 |
| 1000 | 1000 | 0.987 | 0.0008 | 0.990 | 0.0008 | 0.987 | 0.0017 | 0.989 | 0.0007 |
| 1000 | 2500 | 0.988 | 0.0005 | 0.989 | 0.0004 | 0.977 | 0.0065 | 0.989 | 0.0004 |
| 1000 | 5000 | 0.989 | 0.0003 | 0.989 | 0.0003 | 0.980 | 0.0058 | 0.989 | 0.0003 |

In S2, we observe that the GMM clustering is improved at every $m$ and $n$, when compared to S1, due to the added information from the correlation between the B-spline bases. Here, we do see that there is a small loss in performance from using a $k$-means approach over GMM clustering. However, we again observe that there is little effect on clustering performance when the trimmed $k$-means algorithm is used (even when the sample size is halved), rather than the full data $k$-means, other than an increase in the variability of the ARIs. Considering the significant gains in computational speed and the appearance of only a minor loss in performance from using a trimmed $k$-means approach over a full data $k$-means or GMM clustering, and considering the scaling in accuracy due to increasing $m$ and $n$, we find our overall approach justifiable for large data sets such as our calcium imaging application.

Table 2: Average time and ARI results for clustering rules applied to S1. The columns denoted $\alpha = 0.25$ and $\alpha = 0.5$ contain the results for the trimmed $k$-means estimated clusterings. ARI results are reported as means over 50 repetitions.

| | | GMM | | $k$-means | | $\alpha = 0.25$ | | $\alpha = 0.5$ | |
|---|---|---|---|---|---|---|---|---|---|
| $m$ | $n$ | ARI | SE | ARI | SE | ARI | SE | ARI | SE |
| 100 | 500 | 0.976 | 0.0015 | 0.932 | 0.0028 | 0.917 | 0.0052 | 0.910 | 0.0103 |
| 100 | 1000 | 0.981 | 0.0011 | 0.933 | 0.0020 | 0.928 | 0.0040 | 0.931 | 0.0022 |
| 100 | 2500 | 0.984 | 0.0005 | 0.934 | 0.0012 | 0.928 | 0.0043 | 0.930 | 0.0020 |
| 100 | 5000 | 0.985 | 0.0003 | 0.934 | 0.0008 | 0.934 | 0.0008 | 0.932 | 0.0008 |
| 200 | 500 | 0.985 | 0.0015 | 0.951 | 0.0019 | 0.943 | 0.0025 | 0.917 | 0.0123 |
| 200 | 1000 | 0.990 | 0.0007 | 0.951 | 0.0016 | 0.946 | 0.0044 | 0.942 | 0.0062 |
| 200 | 2500 | 0.992 | 0.0003 | 0.951 | 0.0012 | 0.942 | 0.0048 | 0.949 | 0.0012 |
| 200 | 5000 | 0.992 | 0.0003 | 0.951 | 0.0006 | 0.950 | 0.0006 | 0.950 | 0.0007 |
| 500 | 500 | 0.990 | 0.0011 | 0.955 | 0.0026 | 0.928 | 0.0095 | 0.939 | 0.0085 |
| 500 | 1000 | 0.994 | 0.0005 | 0.959 | 0.0016 | 0.950 | 0.0052 | 0.943 | 0.0089 |
| 500 | 2500 | 0.996 | 0.0003 | 0.962 | 0.0010 | 0.961 | 0.0010 | 0.961 | 0.0010 |
| 500 | 5000 | 0.995 | 0.0002 | 0.960 | 0.0008 | 0.957 | 0.0026 | 0.959 | 0.0008 |
| 1000 | 500 | 0.993 | 0.0009 | 0.959 | 0.0021 | 0.937 | 0.0100 | 0.945 | 0.0073 |
| 1000 | 1000 | 0.995 | 0.0004 | 0.963 | 0.0014 | 0.958 | 0.0028 | 0.955 | 0.0065 |
| 1000 | 2500 | 0.996 | 0.0003 | 0.963 | 0.0011 | 0.954 | 0.0039 | 0.962 | 0.0011 |
| 1000 | 5000 | 0.996 | 0.0002 | 0.963 | 0.0005 | 0.959 | 0.0030 | 0.955 | 0.0067 |

# 4 Example Application

## 4.1 Data Description

We consider an analysis of a time series data set arising from the volumetric calcium imaging of a larval zebrafish brain. The in-vivo calcium imaging was performed on a 5 day post fertilization *Tg(elavl3:H2B-GCaMP6s)* fish [50]. Images were acquired using SCAPE microscopy with 488 nm excitation [11]. The zebrafish was imaged at rest over approximately 20 minutes and $640 \times 130 \times 300$ $(x \times y \times z)$ voxel volume (actual dimension $\approx 800 \times 600 \times 200$ $\mu$m$^3$) of time series data were acquired at 4 volumes per second.

As an example, we analyze a central region of interest within the brain, with volume $170 \times 70 \times 150$ $(x \times y \times z)$ voxels. Each time point of each time series measures image intensity on a scale between 0 and 1. The total number of time points at each voxel of the data is $m = 1935$; due to irregularities, each series covers the wider time range of $t_1 = 1$ to $t_m = 2797$ after cropping. All time series are observed at the same $m$ time points. The volume contains a total of $n = 1785000$ spatially correlated time series from voxels potentially displaying interesting neuronal activity or anatomical features. A time averaged image of the 75th $z$-slice of the image is presented in Figure 1. Upon inspection of Figure 1, we see that there exist spatial variation in the mean signal that may indicate the presence of voxel subpopulations.

## 4.2 Data Analysis via Two-Stage Approach

All statistical analyses in this article are conducted within the *R* statistical environment [51] via the *Microsoft Open* 3.2.3 build. Data analysis is conducted using a mix of open source packages along with bespoke scripts. All computations are performed on a *MacBook Pro* (retina, 15-inch, early 2013) with a 2.4 GHz *Intel Core i7* processor and 16 GB of DDR3 RAM. Computation times are measured using the *proc.time* function in the *base* package of *R*.

Imaging data were first converted from the raw time frames (*ANDOR* format) using a custom *python* script. Part of this pre-processing involves a conversion to *MINC* [52], a HDF5 based format such that we can more easily deal with datasets that approximate 1TB. Once in the *MINC* format, the data is then realigned within each volume, realigned for each time series, and intensity normalized using the *MINC* tools. This post-processing is performed on a $\approx 200$ core *Linux* cluster using gridengine. A sub-section of the data are then extracted and converted to *NIfTI* format for statistical analysis. The *NIfTI* format images were read into *R* via the *AnalyzeFMRI* package [53]. Before initializing Stage 1 of our process, we firstly detrend the data. This is performed rapidly via the *speedlm.fit* function within the *speedglm* package [54]. Using a $d = 100$ bases (i.e. a 98-node) B-spline system, the OLS estimates of the B-spline representation $\tilde{\boldsymbol{b}}_i$ $(i = 1, ..., 1785000)$ for Stage 1 of the procedure are quickly computed via the functions within the *fda* package [55]. An example sample of 10 time series $\boldsymbol{Z}_i^T = (Z_i(1), ..., Z_i(2797))$ along with their estimated B-spline representations $\tilde{Y}(t) = \tilde{\boldsymbol{b}}_i^T \boldsymbol{x}(t)$ $(t = 1, ..., 2797)$ is visualized in Figure 2. Reading, detrending, and OLS estimation of every voxel in the volume was performed in a total computation time of 0.89 hours.

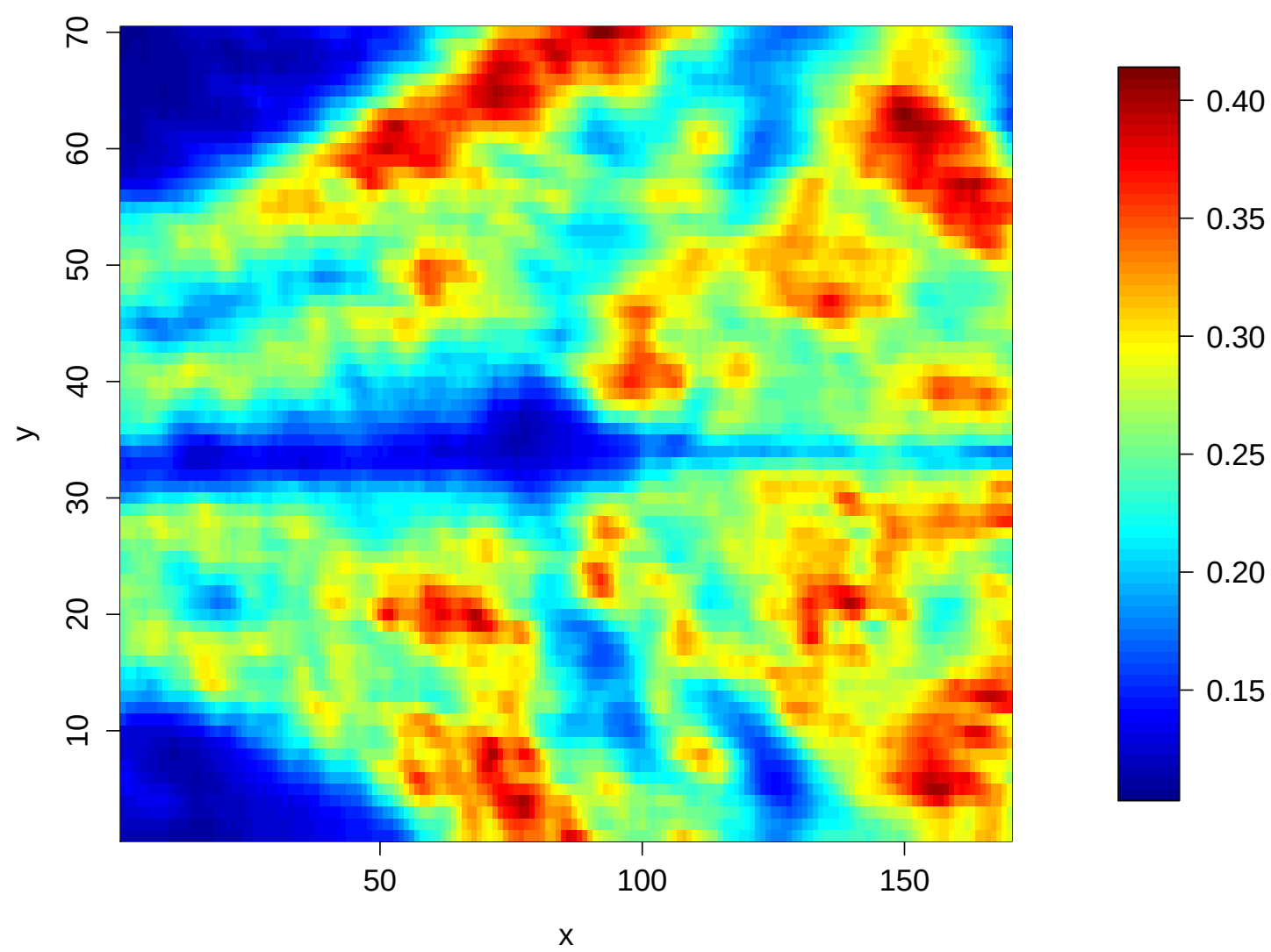

Figure 1: Time-averaged image of the 75th $z$-slice of the zebrafish brain calcium imaging volume acquired using SCAPE microscopy. The legend on the right-hand side indicates the average level of signal intensity at any particular voxel.

Prior to conducting Stage 2 of the process, we firstly column normalize the obtained OLS estimates $\tilde{\boldsymbol{b}}_i^T = \left(\tilde{b}_{i1}, ..., \tilde{b}_{id}\right)$. That is, for each $j = 1, ..., d$, we normalize each of the $j$th column of coefficients $\tilde{b}_{1j}, ..., \tilde{b}_{nj}$ by the respective mean and standard deviation. This is done to reduce the effects of differing dimensional scales. Upon normalization, the trimmed $k$-means algorithm is used to cluster the data. The TCLUST algorithm was applied via the *tkmeans* function from the *tclust* package [56]. Numbers of clusters in the set $\mathbb{K} = \{2, ..., 50\}$ were considered, and the algorithm was repeated 20 times for each $k \in \mathbb{K}$, in order to mitigate against convergence to a spurious solution. The solution that maximized the objective (9) over all repetitions is taken to be the optimal $k$-cluster trimmed $k$-means estimator $\tilde{\boldsymbol{\gamma}}_n^{[k]}$.

Using the objective sequence $\ell_n\left(\tilde{\boldsymbol{\gamma}}_n^{[k]}\right)$ $(k \in \mathbb{K})$, we utilize the *capushe* package [57] to implement the the DDSE method of [29]. Under the penalty $\text{pen}(k) = dk$, The slope heuristic is estimated to be $\tilde{\kappa} = 1.335 \times 10^{-3}$. This

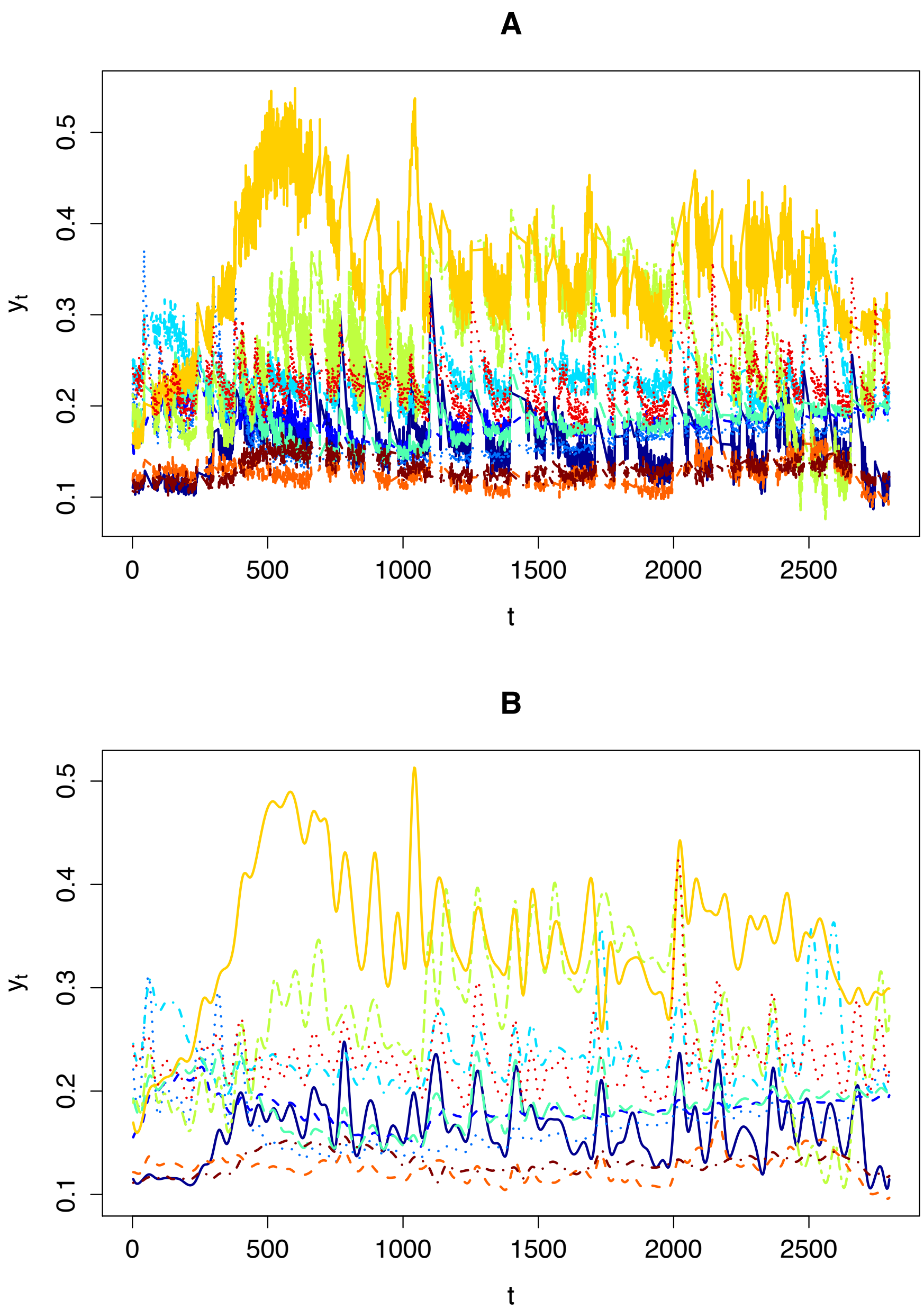

Figure 2: Panel A visualizes 10 randomly sampled time series of intensities from the total of $n = 1785000$ voxels; different colors indicate the signals from different time series. Panel B visualizes the $d = 100$ basis (98-node) B-spline representations $\tilde{Y}(t) = \tilde{\boldsymbol{b}}_i^T \boldsymbol{x}(t)$ for each of the 10 randomly sampled series; the colors correspond to those from Panel A.

results in the model selection rule

$$\hat{k} = \arg\min_{k \in \mathbb{K}} \; -\frac{\sum_{i=1}^{n} \log \sum_{c=1}^{k} \pi_c \phi_d \left(\tilde{\boldsymbol{b}}_i; \tilde{\boldsymbol{\mu}}_{cn}^{[k]}, \mathbf{I}_d\right)}{n} + \frac{2.67kd}{10^3}. \tag{10}$$

Under Rule 10, the optimal number of clusters is determined to be $\hat{k} = 10$. Panel A of Figure 3 visualizes the average time taken to perform 20 repetitions of the TCLUST algorithm for each $k \in \mathbb{K}$; Panel B of Figure 3 visualizes the model selection criterion values $-n^{-1}\ell_n\left(\tilde{\boldsymbol{\gamma}}_n^{[k]}\right) + 2\tilde{\kappa}\,\text{pen}\,(k)$ for each considered $k$. The total computation time for 20 repetitions of the TCLUST algorithm for each $k$ is 82.61 hours.

### 4.3 Results

Upon reversing the normalization discussed in Section 4.2, we can write the $\hat{k} = 10$ cluster mean functions as $\tilde{\mu}_c(t) = \tilde{\boldsymbol{\mu}}_{cn}^T \boldsymbol{x}(t)$ $(c = 1, ..., 10)$. A plot of the 10 cluster mean functions appears in Panel A of Figure 4; a frequency plot of the number of voxels that are allocated into each cluster (via Rule (8)) appears in Panel B of Figure 4. From Figure 4, we notice that although the majority of mean functions appear parallel, there are a few that exhibit behavior that differs over time, and not simply differing in mean; for example, the $c = 3$ and $c = 9$ mean functions exhibit behavior that is substantially different to the others.

A visualization of the clustering at the 75th slice is presented in Figure 5. From Figure 5, we can make some inferences regarding the nature of the voxels that are allocated to each cluster. For example, $c = 1$ can be inferred as background or non-brain matter. Clusters $c = 2, 3, 4, 5, 6, 7$ can be inferred as edge effects on the interface between the background and the brain matter; the differences in these background effects may be explained by the various types of cellular materials such as membranes and tissues. Clusters $c = 8, 9, 10$ can be inferred as various types of brain matter. Furthermore, the cluster allocations in Figure 5 appears to be spatially coherent, thus indicating that the methodology is producing biologically meaningful results and not spurious allocations. Further, we observe that the frequencies of the different clusters are quite varied, indicating that the process is able to identify both rare and common subpopulations of voxels.

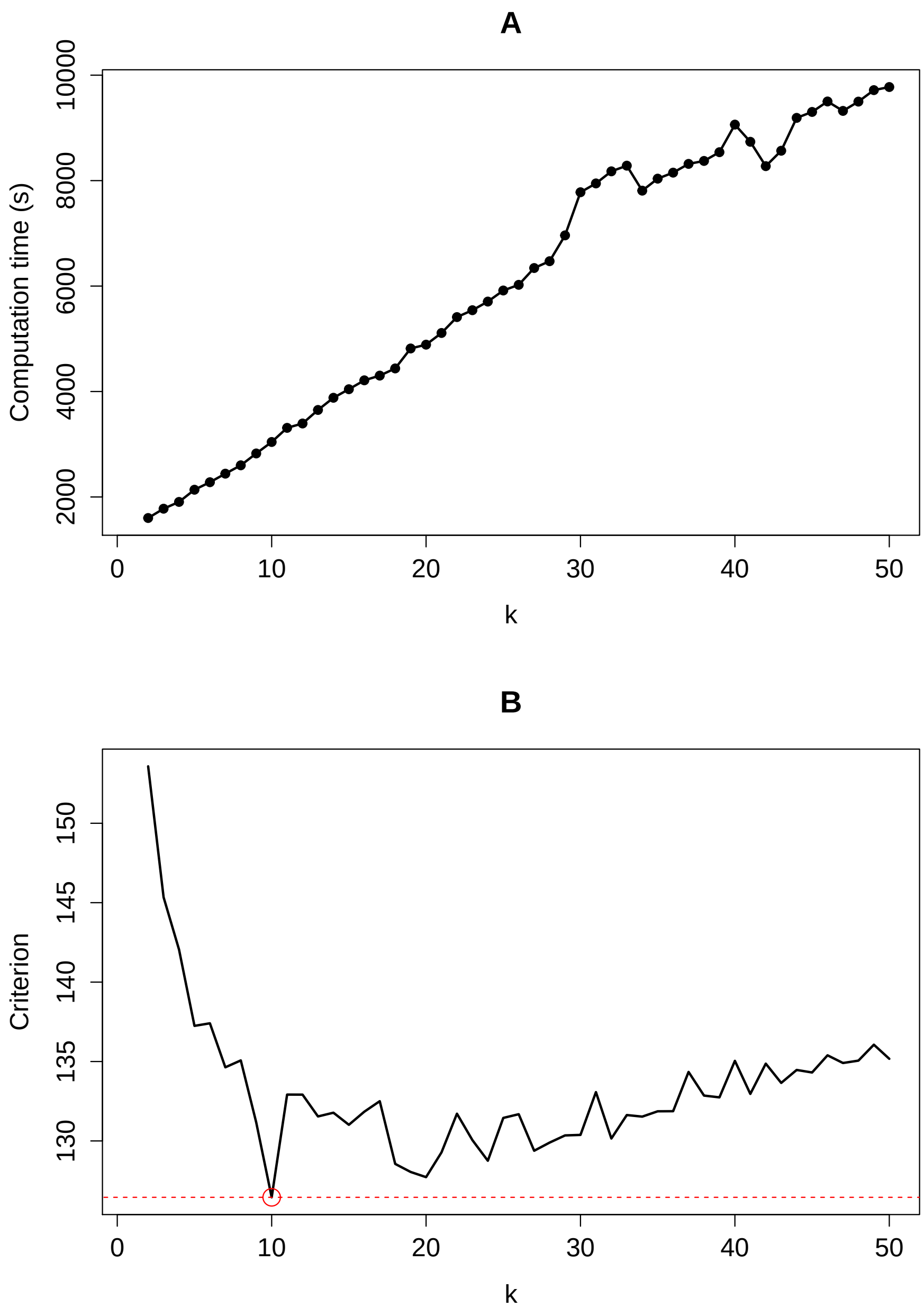

Figure 3: Panel A visualizes the total amount of time (in seconds) taken to run the TCLUST algorithm for each $k \in \mathbb{K}$ ($\mathbb{K} = \{2, ..., 50\}$). Panel B visualizes the model selection criterion values for each $k$, as per Rule (10); the dotted line and circled point indicate the optimal value of the criterion and $\hat{k} = 10$, respectively.

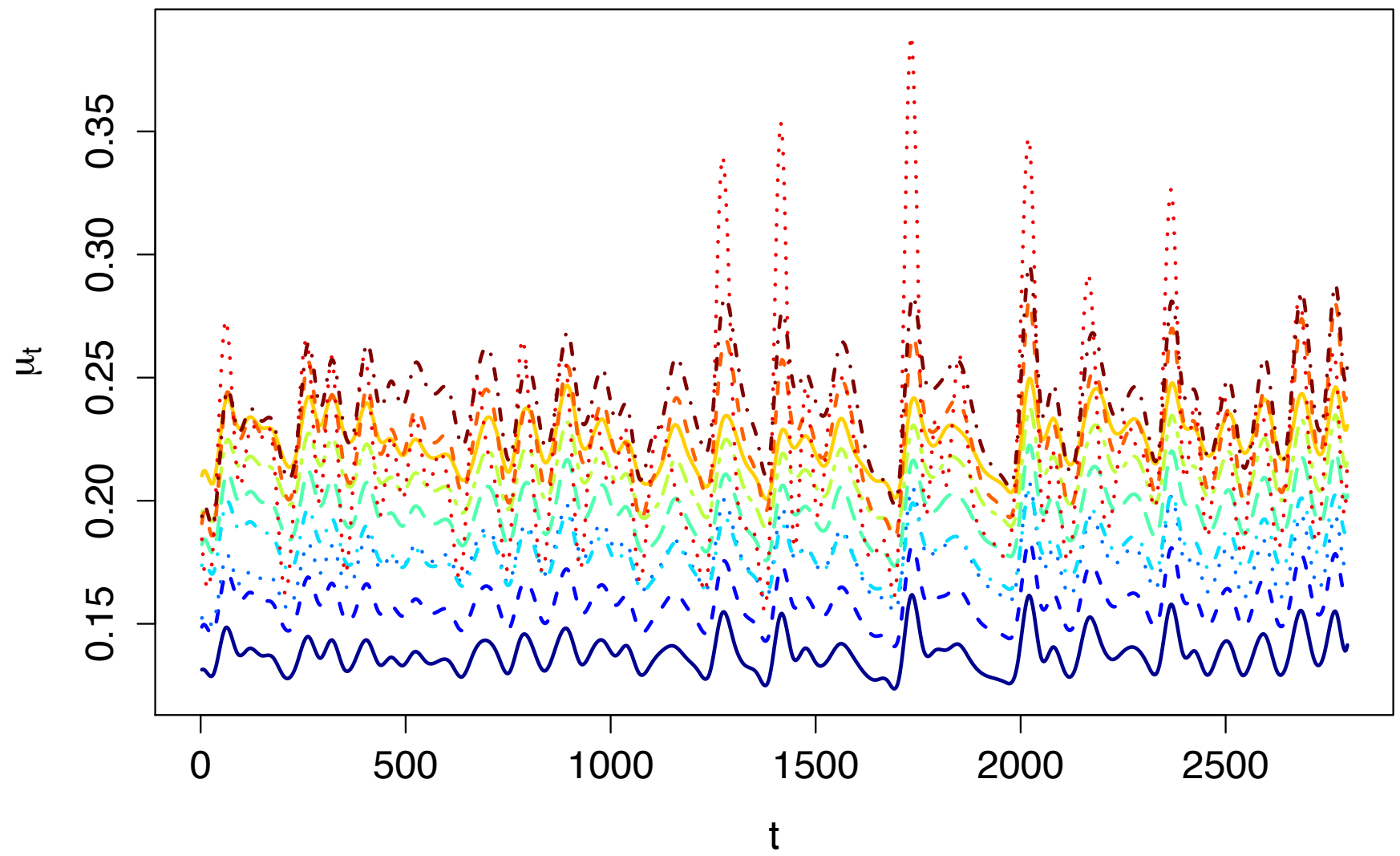

Figure 4: Visualization of the $\hat{k} = 10$ mean functions $\tilde{\mu}_c(t) = \tilde{\boldsymbol{\mu}}_{cn}^T \boldsymbol{x}(t)$ ($c = 1, ..., 10$) that are obtained via the TCLUST algorithm for trimmed $k$-means; the colors are as in Figure 5.

# 5 Conclusions

High-speed volumetric microscopy technologies hold great potential for the investigation of neurological activity in animals; see for example [11]. The methodology for analysis of whole-volume calcium imaging data has not been well developed and especially not adapted to the Big Data pathologies that are inherit in the class of data. In this article, we have developed a model-based clustering method that addresses the problems of sparse sampling, high dimensionality, and spatial dependency, which are inherent in whole-volume calcium imaging data. Our methodology is a two-stage mixture model-based approach for the functional data analysis of time series that arise from calcium imaging experiments.

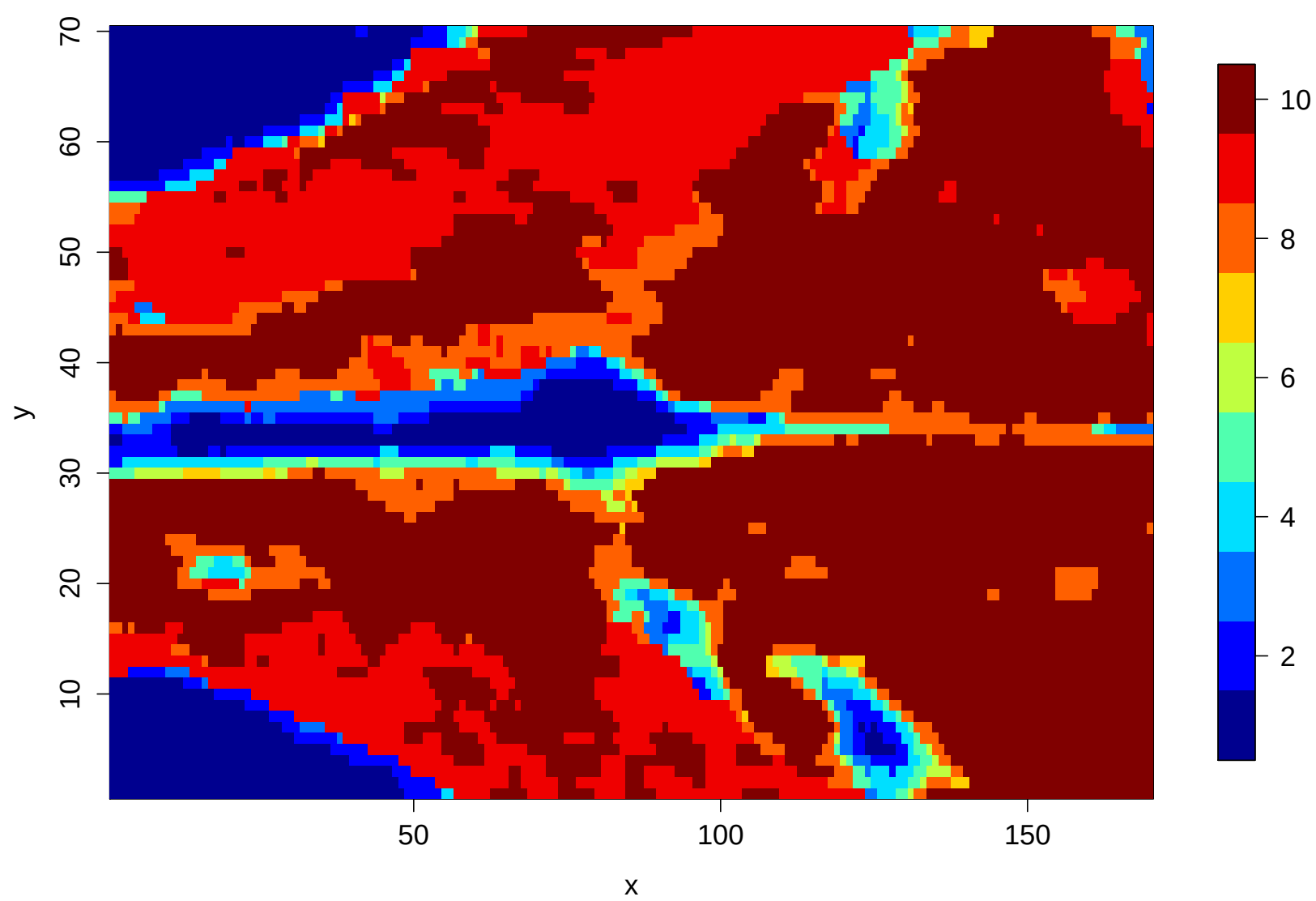

Figure 5: Cluster allocations (as per Rule (8)) of the 75th $z$-slice of the zebrafish brain calcium imaging volume. The legend on the right-hand side indicates the cluster allocation and corresponds with the colors that are used in Figure 4.

We derived a theoretical model under which our approach can be interpreted. The approach is demonstrated to coincide with the estimation of a GMM for OLS estimates of B-spline coefficients. The consistency (under data dependence) of the MLE of the model parameter that is required for the application of the approach is provided. For feasibility, a a simplification of the GMM is estimated using a trimmed $k$-means algorithm. A consistency result (under data dependence) is proven for the trimmed $k$-means estimators. This consistency result is novel and extends the previous results of [30] and [48] regarding the consistency of such estimators under independence of observations.

An example application of this method on a resting-state zebrafish calcium imaging dataset is presented. The computational timing of the approach, using freely-available open software, demonstrated the feasibility of the procedure. The outcome of the data analysis showed that the approach produces spatially coherent cluster allocations that are biologically interpretable and meaningful.

Future directions for this research will involve the application of the approach on animals undergoing stimulation; such data will likely exhibit more functional asychronicities than those observed in the provided example. Computational times may be made faster via bespoke software for the problem, that are implemented in $C$, this may be incorporated in the current workflow via the *Rcpp* package [58]. The bespoke software may also make possible the application of less restrictive models for clustering.

Finally, although we note the spatial coherency of the obtained cluster allocations, such allocations can be made more spatially consistent via the ap-plication of an MRF (Markov random field) model, such as those successfully implemented in [59] and [24]. The MRF model can be applied as a third stage to the approach.

## Acknowledgments

This work was supported by a University of Queensland Early Career Award to JFPU, a NIH BRAIN initiative grant U01 NS094296 and DoD MURI W911NF-12-1-0594 to EH. HDN, GJM, and AJ are supported under the Australian Research Council's Linkage Projects funding scheme (LP130100881). We would also like to thank Misha Ahrens for providing us with zebrafish and Sharon Kim (Hillman lab) for assistance with fish protocols and organization.

## Appendix

### Description of Trimmed $k$-means

Let $\boldsymbol{u}_1, ..., \boldsymbol{u}_n$ be a realization of an ID (identically distributed) random sample $\boldsymbol{U}_1, ..., \boldsymbol{U}_n \in \mathbb{R}^d$. Based on [48, Prop. 1], we can describe the $\alpha$-trimmed $k$-means problem as the estimation of the mean vectors $\boldsymbol{\gamma}^T = (\boldsymbol{\mu}_1, ..., \boldsymbol{\mu}_k)$, where $\boldsymbol{\mu}_c$ $(c = 1, ..., k)$, via the maximization of the objective

$$\mathcal{L}_n(\boldsymbol{\gamma}) = \frac{1}{n} \sum_{i=1}^{n} \sum_{c=1}^{k} \delta_c(\boldsymbol{u}_i; \boldsymbol{\gamma}) \log D_c(\boldsymbol{u}_i; \boldsymbol{\gamma}), \tag{11}$$

where $D_c(\boldsymbol{u}; \boldsymbol{\gamma}) = k^{-1} \phi_d(\boldsymbol{u}; \boldsymbol{\mu}_c, \lambda \mathbf{I}_d)$ and

$$\begin{aligned} \delta_c(\boldsymbol{u}; \boldsymbol{\gamma}) &= \mathbb{I}\Bigg\{ \boldsymbol{u} \in \Big\{ \max_{c'} D_{c'}(\boldsymbol{u}; \boldsymbol{\gamma}) = D_c(\boldsymbol{u}; \boldsymbol{\gamma}) \Big\} \\ &\quad \cap \{ D_c(\boldsymbol{u}; \boldsymbol{\gamma}) \geq R(\boldsymbol{\gamma}) \} \Bigg\}. \end{aligned} \tag{12}$$

Here $\lambda > 0$ is a fixed constant, $\mathbb{I}(A)$ equals 1 if $A$ is true and 0 otherwise, and $R(\boldsymbol{\gamma}) = \inf_v \{G(v; \boldsymbol{\gamma}) \geq \alpha\}$, where $G(v; \boldsymbol{\gamma}) = \mathbb{P}(\max_c D_c(\boldsymbol{u}; \boldsymbol{\gamma}) \leq v)$ under the distribution of $\boldsymbol{U}_1$. Call the $\alpha$-trimmed $k$-means estimate $\tilde{\boldsymbol{\gamma}}_n^T = (\tilde{\boldsymbol{\mu}}_{n1}, ..., \tilde{\boldsymbol{\mu}}_{nk})$.

Upon inspecting the objective (11), we observe that (12) sets the contribution of all observations $\boldsymbol{u}_i$ that have $\max_c D_c(\boldsymbol{u}_i; \boldsymbol{\gamma})$ values below the $\alpha$-quantile to 0; traditionally, the observations below the $\alpha$-quantile are seen to be outliers. Thus, only $(1-\alpha) \times 100\%$ of the data contributes non-zero values to the computation of the objective. This explains the faster computational speed of the trimmed $k$-means over its untrimmed counterpart, when $\alpha$ is sufficiently large. With respect to the size of data in our application, we found $\alpha = 0.9$ to be reasonable.

## TCLUST Algorithm

Like the $k$-means problem, the $\alpha$-trimmed $k$-means problem is combinatorial and thus difficult to solve exactly. An approximate solution for maximizing (11) is to use the TCLUST algorithm. Here, we adapt the description of the TCLUST algorithm (given in [48, Sec. 3]) for the $\alpha$-trimmed $k$-means problem.

Let $\boldsymbol{\gamma}^{(0)}$ denote some randomized initialization of the vector $\boldsymbol{\gamma}$, and let $\boldsymbol{\gamma}^{(r)T} = \left(\boldsymbol{\mu}_1^{(r)}, ..., \boldsymbol{\mu}_k^{(r)}\right)$ denote the $r$th iterate of the algorithm. At the $(r+1)$ th iterate of the algorithm, perform the steps:

1. Compute $d_i^{(r)} = \max_c D_c\left(\boldsymbol{u}_i; \boldsymbol{\gamma}^{(r)}\right)$ and store the set of $\boldsymbol{u}_i$ with the $\lfloor n(1-\alpha) \rfloor$ largest $d_i^{(r)}$ values in the set $\mathbb{H}^{(r+1)}$.

2. Split $\mathbb{H}^{(r+1)}$ into sets $\mathbb{H}_1^{(r+1)}, ..., \mathbb{H}_k^{(r+1)}$, where

$$\mathbb{H}_c^{(r+1)} = \left\{\boldsymbol{u}_i \in \mathbb{H}^{(r+1)} : D_c\left(\boldsymbol{u}_i; \boldsymbol{\gamma}^{(r)}\right) = d_i^{(r)}\right\}.$$

3. For each $c = 1, ..., k$, compute the updates

$$\boldsymbol{\mu}_c^{(r+1)} = \left|\mathbb{H}_c^{(r+1)}\right|^{-1} \sum_{i=1}^{n} \boldsymbol{u}_i \mathbb{I}\left\{\boldsymbol{u}_i \in \mathbb{H}_c^{(r+1)}\right\}.$$

Here $\lfloor x \rfloor$ is the floor of $x \in \mathbb{R}$. Steps 1–3 are repeated until some convergence criterion is met upon which the final iterate is declared the $\alpha$-trimmed $k$-means estimator $\tilde{\boldsymbol{\gamma}}_n$. In this article, we follow the default option of the *tkmeans* function and terminate the algorithm after 20 iterations.

## Theoretical Considerations

The consistency of the $\alpha$-trimmed $k$-means estimator $\tilde{\boldsymbol{\gamma}}_n$ is established for the case of independent and identically distributed random variables in [48]; see also [30]. Due to the application to CI, we require that the $\alpha$-trimmed $k$-means estimator be consistent for data with some dependence structure. Let $\preceq$ denote the lexicographic order relation in $\mathbb{R}^d$ (cf. [60, Example 2.1.10 (3)]), and let $\Gamma = \left\{\boldsymbol{\gamma} \in \mathbb{R}^{kd} : \boldsymbol{\mu}_1 \preceq ... \preceq \boldsymbol{\mu}_k\right\}$. Further, let $l(\boldsymbol{u};\boldsymbol{\gamma}) = \sum_{c=1}^{k} \delta_c(\boldsymbol{u};\boldsymbol{\gamma}) \log D_c(\boldsymbol{u};\boldsymbol{\gamma})$. The following proposition establishes one such result.

**Proposition 3.** *Let $\boldsymbol{U}_1, ..., \boldsymbol{U}_n \in \mathbb{R}^d$ be an ID and stationary ergodic (or strong-mixing) random sample from a continuous distribution and assume that $|cov(\boldsymbol{U}_1)| < \infty$. Assume that there exists some $\boldsymbol{\gamma}_0 \in \Gamma$ such that for every open subset $\mathcal{G} \subset \Gamma$, $\boldsymbol{\gamma}_0 \in \mathcal{G}$ implies that $\mathbb{E}l(\boldsymbol{U}_1;\boldsymbol{\gamma}_0) > \sup_{\boldsymbol{\gamma}\in\Gamma\backslash\mathcal{G}} \mathbb{E}l(\boldsymbol{U}_1;\boldsymbol{\gamma})$. If $\tilde{\boldsymbol{\gamma}}_n \in \Gamma$ is a sequence of estimators such that $\tilde{\boldsymbol{\gamma}}_n = \arg\max_{\boldsymbol{\gamma}\in\Gamma} \mathcal{L}_n(\boldsymbol{\gamma})$, then $\tilde{\boldsymbol{\gamma}}_n \xrightarrow{P} \boldsymbol{\gamma}_0$.*

*Proof.* We invoke the M-estimator theorem of [61, Thm. 2.12], which requires we establish that

$$\sup_{\boldsymbol{\gamma}\in\Gamma} \left|\mathcal{L}_n(\boldsymbol{\gamma}) - \mathbb{E}l(\boldsymbol{U}_1;\boldsymbol{\gamma})\right| \xrightarrow{P} 0.$$

Under the ID stationary ergodic (or strong-mixing) condition, we can use the ULLN of [38, Thm. 5] by verifying that $\mathbb{E}l(\boldsymbol{U}_1;\boldsymbol{\gamma}) < \infty$ (which also veri-fies the use of an appropriate law of large numbers; e.g. [37, Thm. 3.34]). Since $\delta_c(\boldsymbol{u}_1;\boldsymbol{\gamma})$ are binary, we require that $\mathbb{E}\log D_c(\boldsymbol{u};\boldsymbol{\gamma})$ be finite for each $c$. Since $\log D_c(\boldsymbol{u};\boldsymbol{\gamma})$ is nonlinear in $\boldsymbol{u}$ via only quadratic terms, we have $\mathbb{E}\log D_c(\boldsymbol{u};\boldsymbol{\gamma}) < \infty$ and hence $\mathbb{E}l(\boldsymbol{u};\boldsymbol{\gamma}) < \infty$, under the assumption that $|\text{cov}(\boldsymbol{U}_1)| < \infty$. $\square$

*Remark* 8. The constraint of the parameter space to $\Gamma$ is require to break the symmetries of potential solutions. Without the restriction, the assumption that there exists some $\boldsymbol{\gamma}_0 \in \Gamma$ such that for every open subset $\mathcal{G} \subset \Gamma$, $\boldsymbol{\gamma}_0 \in \mathcal{G}$ implies that $\mathbb{E}l(\boldsymbol{U}_1;\boldsymbol{\gamma}_0) > \sup_{\boldsymbol{\gamma}\in\Gamma\backslash\mathcal{G}} \mathbb{E}l(\boldsymbol{U}_1;\boldsymbol{\gamma})$ cannot hold, since there will always be a $\boldsymbol{\gamma} \in \Gamma\backslash\mathcal{G}$ such that $\mathbb{E}l(\boldsymbol{U}_1;\boldsymbol{\gamma}_0) = \mathbb{E}l(\boldsymbol{U}_1;\boldsymbol{\gamma})$. The lexicographic ordering restriction is similar to that which is used by [62], in order to break symmetries in mixture of experts models.

*Remark* 9. As with Proposition 2, an $M$-dependence assumption can be used to enforce strong-mixing; see Remark 5. However, similarly to Remark 6, $M$-dependence may be difficult to establish if all dimensions of the calcium imaging is allowed to extend infinitely. An alternative theorem based on the results of [41] can be used instead in such a case.